\documentclass[12pt, a4paper]{article}

\usepackage{amsmath}
\usepackage{amssymb}
\usepackage{amsthm}
\usepackage{graphics}

\newtheorem{thm}{Theorem}
\newtheorem{lem}{Lemma}

\newcommand{\adj}[1]{\mbox{$#1^\dag$}}
\newcommand{\ket}[1]{\mbox{$|#1\rangle$}}
\newcommand{\bra}[1]{\mbox{$\langle#1|$}}
\newcommand{\inner}[2]{\mbox{$\langle#1|#2\rangle$}}

\newcommand{\squash}[1]{\raisebox{0.04ex}[0pt][0pt]{\small$\textstyle #1$}}

\title{Positive Operation Valued Measurement Based Multi-User Detection in DS-CDMA Systems}
\author{S\'{a}ndor Imre, Ferenc Bal\'{a}zs\\%}
Budapest University of Technology and Economics\\
Department of Telecommunications\\ Mobile Communications
Laboratory \\ H-1117 Budapest, Magyar Tud\'{o}sok krt. 2., HUNGARY \\
Email: IMRE@hit.bme.hu - BALAZSF@hit.hit.bme.hu}
\date{June 28, 2001}

\begin{document}
\maketitle \abstract {3G and 4G mobile are based on CDMA
technology. In order to increase the effectiveness of CDMA
receivers large amount of effort is invested to develop suitable
multi-user detector techniques. However, at this moment there are
only suboptimal solutions available because of the rather high
complexity of optimal detectors. One of the possible receiver
technologies can be the quantum assisted computing devices which
allows high level parallelism in computation. The first
commercial devices are estimated by 2004, which meets the advert
of 3G and 4G systems. In this paper we introduce a novel quantum
computation based Quantum Multi-user detection (QMUD) algorithm,
employing simple Positive Operation Valued Measurement (POVM),
which provides optimal solution. The proposed algorithm is robust
to any kind of noise.}
\\ \\ \textit{Keywords: Multi-user detection, Positive Operation
Valued Measurement, Quantum computing, Quantum Signal Processing}

\section{Introduction}
The subscribers of next generation wireless systems will
communicate simultaneously, sharing the same frequency band. All
around the world in 3G mobile systems apply Direct Sequence -
Code Division Multiple Access (DS-CDMA) promising due to its high
capacity and inherent resistance to interference, hence it comes
into the limelight in many communication systems. Another
physical layer scheme, Orthogonal Frequency Division Access
(OFDM), is also often used e.g. for Wireless LANs (WLAN) or
HiperLAN, where the subscriber's signal is transmitted via a group
of orthogonal frequencies, providing Inter Channel Interference
(ICI) exemption. Nevertheless due to the frequency selective
property of the channel, in case of CDMA communication the
orthogonality between user codes at the receiver is lost, which
leads to performance degradation. Single-User detectors were
overtaxed and showed rather poor performance even in multi-path
environment \cite{verdu}. To overcome this problem, in recent
years Multi-User Detection (MUD) has received considerable
attention and become one of the most important signal processing
task in wireless communication. \par Verdu \cite{verdu} has
proven that the optimal solution is consistent with the
optimization of a quadratic function, which yields in MLSE
(Maximum-Likelihood Sequence Estimation) receiver. However, to
find the optimum is a \textit{NP}-hard problem as the number of
users grows. Many authors proposed sub-optimal linear and
nonlinear solutions such as Decorrelating Detector, MMSE (Minimum
Mean Square Error) detector, Multistage Detector, Hoppfield
neural network or Stochastic Hoppfield neural network
\cite{verdu,dejou00,var90,aaz92}, and the references therein. One
can find a comparison of the performance of the above mentioned
algorithms in \cite{dejou01}.
\par Nonlinear sub-optimal solutions provide quite good
performance, however, only asymptotically. Quantum computation
based algorithms seem to be able to fill this long-felt gap.
Beside the classical description, which we recently use,
researchers in the early $20^{th}$ century raised the idea of
quantum theory, which nowadays becomes remarkable in coding
theory, information theory and for signal processing
\cite{pres98}. \par Nowadays, every scientist applies classical
computation, using sequential computers. Taking into account that
Moore's law can not be held for the next ten years because
silicon chip transistors reach atomic scale, therefore new
technology is required. Intel, IBM and other companies invest
large amount of research to develop devices based on quantum
principle. Successful experiments share up that within 3-4 years
quantum computation (QC) assisted devices will be available on
the market as enabling technology for 3G and 4G systems. \par
This paper is organized as follows: in Section \ref{sec:0} we
shortly review the necessity of multi-user detection, whereas in
Section \ref{sec:1} the applied quantum computation method is
shown. In Section \ref{sec:2} we discuss the used system model.
In Section \ref{sec:3} the novel MUD algorithm is introduced and
finally we conclude our paper in Section \ref{sec:4}.
\section{Multi-User Detection}\label{sec:0}
One of the major attributes of CDMA systems is the multiple usage
of a common frequency and time slot. Despite the interference
caused by the multiple access property, the users can be
distinguished by their codes. Let us investigate a DS-DCDMA
system, where the $z^{th}$ symbol of the $k^{th}$
$(k=1,2,\dots,K)$ user is denoted by $b_k(z)$. Applying BPSK
modulation, the output signal of the $k^{th}$ user, denoted by
$q_k(t)$, is given as
\[q_k(t)=\sqrt{E_k}\sum_{z=-L}^Lb_k(i)s_k(t-zT),\]
where $s_k(t)$ and $E_k$ is the continuous signature signal and
energy associated to the $k^{th}$ user, $T$ is the time period of
one symbol, and $(2L+1)$ is the size of a block, respectively. For
the sake of simplicity we assume one path propagation channel, so
the channel distortion for the $k^{th}$ user is modeled by a
simple attenuation factor $h_k(t)=a_k$. This model, however, can
be easily applied to more sophisticated channel models, as well.
\par The received signal is the sum of arriving signals plus a
Gaussian noise component and thus can be written as follows:
\begin{eqnarray}
r(t)=\sum_{k=1}^Kh_k(t)\ast q_k(t)+n(t)= \nonumber\\
=\sum_{k=1}^K\sum_{l=-L}^L\sqrt{E_k}a_kb_k(l)s_k(t-lT)+n(t),
\label{eq:mud1}
\end{eqnarray}
where $K$ is the number of users using the same band, $n(t)$ is a
white Gaussian noise with a constant $N_0$ spectral density. In
case of signatures limited to one symbol length and if the system
is properly synchronized, then there is no intersymbol
interference. Consequently, index $z$ can then be omitted from
(\ref{eq:mud1}), yielding:
\[ r(t)=\sum_{k=1}^K\sqrt{E_k}a_kb_ks_k(t)+n(t), t\in [0,T). \]
\par A conventional detector contains $k=1,\dots,K$ filters which
are matched to the corresponding signature waveforms and channels
and calculates the following decision variable:
\begin{equation}
\tilde{b}_k=\sqrt{E_k}a_k\int_0^Tr(t)s_k(t)\mathrm{d}t.
\label{eq:mud2}
\end{equation}
The traditional "single-user" detector (SUD) simply calculate the
sign of expression (\ref{eq:mud2}) yielding
$\hat{b}_k^{SUD}=\mathsf{sign}\{\tilde{b}_k\}$. This method
results in poor performance, as $\tilde{b}_k$ contains not only
the signal transmitted by the $k^{th}$ user but an interference
term generated by the other users:
\[\tilde{b}_k=\underbrace{b_k\varrho_{kk}}_{\mbox{signal}}+\underbrace{\sum_{l=1, l\neq k}^Kb_l\varrho_{kl}}_{\mbox{multiple access interference}}+\underbrace{n_k}_{\mbox{noise}},\]
where $\varrho_{kl}$ is defined as follows:
\begin{equation}
\varrho_{kl}=\sqrt{E_k}\sqrt{E_l}\alpha_k\alpha_l\int_0^Ts_k(t)s_l(t)\mathrm{d}t,
\label{eq:mud3}
\end{equation}
and $n_k=\int_0^Ts_k(t)n(t)\mathrm{d}t$ is a zero mean white
Gaussian noise due to linear transformation. The output of the
matched filter in vector form is:
\begin{equation}
\mathbf{\tilde{b}}=\mathbf{R}\mathbf{b}+\mathbf{n},
\label{eq:mud4}
\end{equation}
where
$\mathbf{\tilde{b}}=[\tilde{b}_1,\tilde{b}_2,\dots,\tilde{b}_K]^T$,
$\mathbf{b}=[b_1,b_2,\dots,b_K]^T$ and
$\mathbf{n}=[n_1,n_2,\dots,n_K]^T$, whereas
$\mathbf{R}=[\varrho_{kl},$ $k=1,2,\dots,K$ $l=1,2,\dots,K]$ is a
symmetric matrix with property of diagonal dominance
$(\varrho_{zz}>\varrho_{zj}$ $\forall j=1,2,\dots,K$ $j\neq i)$.
\par Based on the model introduced above, we derive now the
optimal MUD. The MUD algorithm processes vector
$\mathbf{\tilde{b}}$, which is the output of the matched filter.
To obtain optimal solution based on Bayesian decision rule one
wants to chose the maximal probability binary sequence
conditioned by the received data series. The optimal Bayesian
detection reduces to the following minimization problem
\cite{verdu}: \begin{equation}
\mathbf{\tilde{b}}^{opt}=\min_{\mathbf{y}\in
\{-1,+1\}^K}\left[(\mathbf{\tilde{b}}-\mathbf{R}\mathbf{y})^T\mathbf{R}^{-1}(\mathbf{\tilde{b}}-\mathbf{R}\mathbf{y})\right].
\label{eq:mud5}\end{equation} \par Unfortunately, the search for
the global optimum of (\ref{eq:mud5}) usually proves to be rather
tiresome, which prevents real time detection (its complexity by
exhaustive search is $\mathcal{O}(2^K)$). Therefore, our
objective is to develop new, powerful detection technique, which
paves the way toward real time MUD even in highly loaded system.
\section{Quantum Computation Theory}\label{sec:1}
\par Quantum theory is a mathematical model of a physical system. To
describe such a model we need to specify the representation of a
system. Every physical system can be characterized by means of its
states in the \textit{Hilbert} vector space over the complex
numbers $\mathbb{C}$. The vectors will be denoted as
$\ket{\varphi}$\footnote{Say ket $\varphi$.}. The inner product
$\mbox{\inner{\psi}{\varphi}}$ maps the ordered pair of vectors
to $\mathbb{C}$ with the properties \cite{pres98}
\begin{itemize}
\item Positivity: $\mbox{\inner{\psi}{\psi}}$$>0$ for $\ket{\psi}=0$,
\item Linearity:
$\mbox{\bra{\varphi}(a\ket{\psi_1}+b\ket{\psi_2})=a\inner{\varphi}{\psi_1}+b\inner{\varphi}{\psi_2}}$,
\item Skew symmetry:
$\inner{\varphi}{\psi}=\inner{\varphi}{\psi}^*$.
\end{itemize}
\par In the classical information theory the smallest conveying information unit
 is the \textit{bit}. The counterpart unit
in quantum information is called the \textit{"quantum bit"} the
qubit. Its state can be described by means of the state
$\ket{\varphi}$, $\varphi=\alpha\ket{0}+\beta\ket{1}$, where
$\alpha,\beta \in \mathbb{C}$ refers to the complex probability
amplitudes and $|\alpha|^2+|\beta|^2=1$
\cite{shor98,deutsch00,pres98}. The expression $|\alpha|^2$
denotes the probability that after measuring the qubit it can be
found in $\ket{0}$ computational base, and $|\beta|^2$ shows the
probability to be in computational base $\ket{1}$. In more
general description an $N$-bit qregister $\ket{\varphi}$ is set up
from qbits spanned by $\ket{i}$ $i=0\dots(M-1)$ computational
basis, where $M=2^N$ states can be stored in the same time
\cite{imr01}.
\begin{eqnarray}
\ket{\varphi}=\sum_{i=0}^{M-1}\varphi_i\ket{i} & \varphi_i\in
\mathcal{C}, \label{eq:6}\end{eqnarray} where $M$ denotes the
number of states and $\forall i\neq j \inner{i}{j}=0$,
$\inner{i}{i}=1$, $\sum|\varphi_i|^2=1$, respectively. It is
worth mention, that a $U$ transformation on a qregister is
executed parallel on all $M$ stored states, which is called
\textit{quantum parallelization}. To the irreversiblity of
transformation, $U$ must be unity $U^{-1}=U^T$. The quantum
registers can be set in a general state using quantum gates which
can be represented by means of a unitary operation, described by
a quadratic matrix. Applying four basic gates any states can be
prepared \cite{shor98}.

%\unitlength 1mm
%\begin{picture}(50,20)(-40,15)
%\put(10,24){\line(1,0){10}}
%\put(20,20){\framebox(8,8){H}}
%\put(28,24){\line(1,0){10}}
%\put(38,24){\circle*{1}}
%\put(38,24){\vector(0,-1){9}}
%\put(10,14){\vector(1,0){27}}
%\put(38,14){\circle{2}}
%\put(38.1,14){\makebox(0,0){+}}
%\put(38,24){\line(1,0){10}}
%\put(39,14){\line(1,0){9}}
%\put(12,27){\makebox(0,0){\tiny{$\ket{0}$}}}
%\put(40,27){\makebox(0,0){\tiny{$\oosrt\left(\ket{0}+\ket{1}\right)$}}}
%\end{picture}
\section{System Model}\label{sec:2}
\par In DS-CDMA systems an information bearing bit is encoded by
means of a user specific code with length of the processing gain
(\textit{PG})\cite{verdu}. In case of uplink communication we
assume perfect power control. In the receiver side it is not
required synchronization between input signals and user specific
codes, however, chip synchronization is necessary, that allows
using multipath propagation channels. \par Since classical
multi-user detection schemes only try to minimize the probability
of error in noisy and high interference environment, they, even
also optimal solutions, can commit an error. Actually, these
classical approaches make compromise between computational
complexity, probability of error and time barrier required for
efficient working. On the other hand, QMUD does not make error in
detection, at most QMUD can not make a decision in certain cases,
furthermore, it indicate us this symbol in order to correct in a
higher layer.
\subsection{Representation of Possible Received Sequences in
Qregisters} We quantize every chip of the $k^{th}$ user's
codeword in a qregister of length $N_{ch}$, where the number
representation is not significant at the evaluation of the
received symbol, because with increasing the number of the
alphabet, the precision of number notation grows exponentially. In
our model we prepare for user $k$ two quantum register
$\mbox{\ket{\varphi_1^k}}$ and $\mbox{\ket{\varphi_0^k}}$ each
corresponding to transmitted bit "1" and "0" with an overall
length $N_Q=N_{ch}\cdot PG$. It is important to notice that the
effects of a multi-path channel and the additive noise are
contained in the registers, moreover, the density function of the
noise does not need to be known \textit{a-priori}, however, the
knowledge of noise can reduce the number of qubits in a
qregister. This uncertainty may not influence the exact decision.
Let $V$ denotes a vector space spanned by $\mbox{\ket{v_i}}$,
$i=1\ldots2^{N_Q}$ orthonormal computational base states, where
$\mbox{\inner{v_i}{v_j}=0}$ for $\forall i\neq j$ and
$\mbox{\inner{v_i}{v_j}=1}$ for $\forall i=j$ is hold. The number
of stored states in quantum registers $\mbox{\ket{\varphi_1^k}}$
or $\mbox{\ket{\varphi_0^k}}$ is denoted with $N_{s1}$ or
$N_{s0}$, respectively. If the register
$\mbox{\ket{\varphi_1^k}}$ contains the desired state
$\mbox{\ket{v_i}}$, then
\begin{equation}
\varphi_1^k(i)\equiv \inner{\varphi_1^k}{v_i}= \left\{
\begin{array}{cl}
\mbox{\squash{\frac{1}{\sqrt{N_{s1}}}}} & \mbox{if
$\ket{v_i}\in\ket{\varphi_1^k} \equiv a_i\neq 0$}
\\ 0 & \mbox{otherwise},
\end{array}\right.
\end{equation}
that fulfills the stipulation
$\sum_{i=1}^{2^{N_Q}}\left|\varphi_1^k(i)\right|^2=1$.
\subsection{Preparation of quantum register states}
Due to the effect of multi-path propagation it is required to
form any delayed version of chip sequences of user $k$. This
operation can be made via the so called swap gate, which changes
the position of two qubits in a register. In general, it can be
seen as a quantum shift register. One can think, all the possible
states should be computed before doing quantum multi-user
detection. It is true, however, using classical sequential
computers, this operation could take rather long time, whereas
quantum computation exploits the quantum parallelism. Applying
this feature a transformation on $N$ states stored in a register
can be done in one single step, that provides fast, efficient
preparation of $\mbox{\ket{\varphi_1^k}}$ and
$\mbox{\ket{\varphi_0^k}}$.
\section{Quantum Multi-User Detector}\label{sec:3}
The decision rule of classical multi-user detector becomes a
measurement in quantum world. In our case we have to find out
that the received and quantized signal vector of user $k$
$\mbox{\ket{r^k}=\ket{v_i}}$ is either in the register
$\mbox{\ket{\varphi_1^k}}$ or $\mbox{\ket{\varphi_0^k}}$ or both.
In more mathematical description
\begin{equation}
\begin{array}{ccc}
\inner{\varphi_1^k}{r^k}\stackrel{\textrm{?}}{=}0 & \textnormal{
i.e.} & \varphi_1^k(i)\stackrel{\textrm{?}}{=}0.
\end{array}
\end{equation}
Because of the multi-path propagation and the noise the same state
$\ket{v_i}$ could be found in both registers that makes the
detection impossible. It shall be emphasized, however, that QMUD
is able to recognize this event allowing higher layer protocols
to perform error correction, hence it will never made false
decision, as classical MUD algorithms (independently whether it
is suboptimal or optimal) may do. On the other hand this can not
be seen as feebleness of QMUD since the classical MUD is also
unable to make proper decision in such a situation. The decision
rules of QMUD are showed in Table \ref{tab:1}. From now onward we
only focus on $\mbox{\ket{\varphi_1^k}}$, the operations
on $\mbox{\ket{\varphi_0^k}}$ are analogous.%\vspace{-2mm}
\begin{table}[hbt]\caption{QMUD decision rule table} \label{tab:1}
\begin{center}
\begin{tabular}[t]{|c|c|c|} \hline
$\inner{\varphi_1^k}{r^k}$ & $\inner{\varphi_0^k}{r^k}$ & decision
\\\hline 0 & 0 & no message was sent \\ 0 & $\neq 0$ & the bit "0" was sent\\
$\neq 0 $ & 0 & the bit "1" was sent \\ $\neq 0$ & $\neq 0$ & no
decision is possible \\\hline
\end{tabular}
\end{center}
\end{table}
\vspace{-2mm}
\subsection{Evaluation of $\inner{\varphi_1^k}{r^k}$ - The measurement}
The evaluation of $\inner{\varphi_1^k}{r^k}$ is not a trivial
task as this is not an unitary operation, as discussed in Section
\ref{sec:1}. In the register $\ket{\varphi_1^k}$ there is only one
state $\ket{r^k}=\ket{v_i}$ we are interested in. However, from
measurement point of view the overall state of the quantum
register being is state $\ket{\varphi_1^k}$ can be regarded as a
qubit. This qubit can be written as $\alpha
\ket{0}+\inner{\varphi_1^k}{r^k}\ket{1} $, where
$\alpha=\sqrt{\sum_{j=1,j\neq
i}^{2^{N_Q}}\left|\varphi_1^k(j)\right|^2}\equiv\inner{\varphi_1^k}{v_j}$.
This qubit contains two states $\ket{\eta_1}=\ket{0}$ and
$\ket{\eta_2}=\sqrt{\frac{N_{s1}-1}{N_{s1}}}\ket{0}+\sqrt{\frac{1}{N_{s1}}}\ket{1}$
corresponding to the probability amplitude of $\ket{v_i}$ is in
the register or not. It can be simply proved that $\ket{\eta_1}$
and $\ket{\eta_2}$ are not unambiguously distinguishable, because
$\inner{\eta_1}{\eta_2}\neq 0$. \par However, we can extend the
computational bases and apply the so called Positive Operation
Valued Measurement (POVM-see Appendix). We introduce three
positive operators
\begin{equation}
\mathbf{E_1}=\alpha\ket{1}\bra{1}=\begin{pmatrix} 0 & 0 \\ 0 &
\alpha \end{pmatrix}, \label{eq:3}
\end{equation}
\begin{equation}
\begin{gathered}
\mathbf{E_2}=\beta\left[\sqrt{\frac{1}{N_{s1}}}\ket{0}-\sqrt{1-\frac{1}{N_{s1}}}\ket{1}\right]\left[\sqrt{\frac{1}{N_{s1}}}\bra{0}-\sqrt{1-\frac{1}{N_{s1}}}\bra{1}\right]=
\\=\begin{pmatrix} \beta\frac{1}{N_{s1}} &
-\beta\sqrt{\frac{N_{s1}-1}{N_{s1}^2}} \\
-\beta\sqrt{\frac{N_{s1}-1}{N_{s1}^2}} &
\beta\left(1-\frac{1}{N_{s1}}\right)
\end{pmatrix}\textnormal{      and} \label{eq:4}
\end{gathered}
\end{equation}
\begin{equation}
\mathbf{E_3}=\mathbf{I}-\mathbf{E_1}-\mathbf{E_2}=\begin{pmatrix}
1-\beta\frac{1}{N_{s1}}& \beta\sqrt{\frac{N_{s1}-1}{N_{s1}^2}}
\\ \beta\sqrt{\frac{N_{s1}-1}{N_{s1}^2}} & 1-\alpha-\beta\left(1-\frac{1}{N_{s1}}\right) \end{pmatrix} , \label{eq:5}
\end{equation}
where $\mathbf{I}$ is the identity matrix. The operator
(\ref{eq:5}) provides
\begin{equation}
\sum_{j=1}^3p(E_j)\mid_{\ket{\eta_1}}=\sum_{j=1}^3p(E_j)|_{\ket{\eta_2}}=1,
\end{equation}
besides the first two POVM measurement operators in (\ref{eq:3},
\ref{eq:4}) are orthogonal to $\ket{\eta_1}$ and $\ket{\eta_2}$,
respectively, making the probabilities of measuring $\mathbf{E_1}$
and $\mathbf{E_2}$
\begin{equation}
\begin{gathered}
P(\mathbf{E_1})|_{\ket{\eta_1}}=\bra{\eta_1}\mathbf{E_1}\ket{\eta_1}=0,
\\P(\mathbf{E_2})|_{\ket{\eta_2}}=\bra{\eta_2}\mathbf{E_2}\ket{\eta_2}=0,
\label{eq:13}
\end{gathered}
\end{equation}
where $P(\mathbf{E}_i)|_{\ket{\eta_j}}$ refers to the probability
of the event the $\mathbf{E}_i$ was measured if $\ket{\eta_j}$ had
been received. In other words, if our instrument indicates
$\mathbf{E_1}$, only the information corresponding to the state
$\eta_2$ could be sent, otherwise if $\mathbf{E_2}$ is indicated
the received state must be $\eta_1$. It is appreciable that the
scale of uncertainty arising from POVM measurement is a function
of $\mathbf{E_3}$. It is important to emphasize that detecting
$\mathbf{E_3}$ we do not any false detection. To reduce this
effect the free variables $\alpha$ and $\beta$ in $\mathbf{E_3}$
should be set to zero which makes it to identity matrix.
Unfortunately, in that case, the resulting matrix becomes to a
non-positive definite one.
\subsection{Setting the variables $\alpha$ and $\beta$}
The operator $\mathbf{E}$ is positive if
$\bra{\varphi}\mathbf{E}\ket{\varphi}\geq 0$ for any
$\ket{\varphi}$. A positive definite matrix has the form
$\mathbf{E_3}=(\ket{\mathcal{A},\mathcal{B}}\bra{\mathcal{A},\mathcal{B}})$,
where in our case $\mathcal{A}=\sqrt{1-\beta\frac{1}{N_{s1}}}$ and
$\mathcal{B}=\sqrt{1-\alpha-\beta\left(1-\frac{1}{N_{s1}}\right)}$,
moreover, according to (\ref{eq:5}) the product should satisfy
\begin{equation}
\mathcal{AB}=\sqrt{1-\beta\frac{1}{N_{s1}}}\cdot\sqrt{1-\alpha-\beta\left(1-\frac{1}{N_{s1}}\right)}\stackrel{\textrm{!}}{=}\beta\sqrt{\frac{N_{s1}-1}{N_{s1}^2}},
\label{eq:7}
\end{equation}
which leads to
\begin{equation}
\alpha=\frac{1-\beta}{1-\frac{\beta}{N_{s1}}}, \label{eq:8}
\end{equation}
that makes $\mathbf{E_3}$ positive.
\par We assume at the moment the symbols "1" and "0" are
transmitted with equal probabilities, therefore it is worth
choosing the measurement probabilities
$P(\mathbf{E_1})|_{\ket{\eta_2}}$ and
$P(\mathbf{E_2})|_{\ket{\eta_1}}$ to be equal.
\begin{lem} If the
probabilities measurement
$P(\mathbf{E_1})|_{\ket{\eta_2}}=P(\mathbf{E_2})|_{\ket{\eta_1}}$
then $\alpha=\beta$ furthermore $\alpha=\frac{1}{2}$.
\end{lem}
\begin{proof}
\begin{equation}
\begin{gathered}
P(\mathbf{E_1})|_{\ket{\eta_2}}=\bra{\eta_2}\mathbf{E_1}\ket{\eta_2}=\frac{\beta}{N_{s1}},
\\ \textnormal{and} \\
P(\mathbf{E_2})|_{\ket{\eta_1}}=\bra{\eta_1}\mathbf{E_2}\ket{\eta_1}=\frac{\alpha}{N_{s1}}
\label{eq:9}
\end{gathered}
\end{equation}
Substituting $\alpha=\beta$ in (\ref{eq:8}) one gets a quadratic
function with roots of
$\alpha_1=N_{s1}-\sqrt{N_{s1}\left(N_{s1}-1\right)}$ and
$\alpha_2=2N_{s1}$, where the latter one is impossible since
probability can not become greater than 1. Moreover, $\alpha_1$
converges very fast to $\frac{1}{2}$ as $N_{s1}$ goes to infinity.
\end{proof}
However, as the length $N_{s1}$ of a register grows the resulting
probability of detection $\frac{\beta}{N_{s1}}$ becomes always
smaller, due to the small angle between $\ket{\eta_1}$ and
$\ket{\eta_2}$. POVM \cite{grun97} is typically used in such
situations, where the measurement can not be repeated. In our
case, however, in our system the content of the registers
$\mbox{\ket{\varphi_1^k}}$ and $\mbox{\ket{\varphi_0^k}}$ are
constant during detection, allowing multiple measurements or even
parallelization of them, which makes $P(\mathbf{E_3})$ smaller at
every step. \par Although, lemma 1. shows that $\alpha=\beta$ is
an expedient choice, with additional thoughts $P(\mathbf{E}_3)$
can be further reduce. Therefore, we assert the next theorem.
%\newpage
\begin{thm}
Using appropriate different values for $\alpha$ and $\beta$ one
can double the probability of detection.
\end{thm}
\begin{proof}
We apply two measurements parallel, where
\[\max_{\beta}P(\mathbf{E_1})|_{\ket{\eta_2}}\Longrightarrow\max_{\beta}\alpha\]
and \[\max_{\beta}P(\mathbf{E_2})|_{\ket{\eta_1}}.\] Focusing
again on the former case, of course, $P(\mathbf{E_2})|_{\eta_1}$
and also $P(\mathbf{E_3})$ become very small. Since the bounds of
a probability variable $x$ must satisfy $0<P(x)<1$, so the bounds
of $\alpha$, $0<\alpha<N_{s1}$ are known, as well. Issued from
(\ref{eq:8}) $\alpha$ takes negative values if $\beta$ becomes
greater than $1$, and the same is held for the opposite case,
respectively. The possible values of $\alpha$ and $\beta$ are
depicted in Figure \ref{pic:1}, where two linear function of
$\beta$ can be seen according to the
numerator and denominator of (\ref{eq:8}).\\
\begin{figure}[h]
\begin{picture}(30,50)(-230,-5)
\unitlength 1mm \thicklines \put(-5,0){\line(1,0){5}} \thinlines
\put(0,0){\line(1,0){8}} \thicklines \put(8,0){\vector(1,0){17}}
\thinlines \put(0,-2){\vector(0,1){15}}
\put(23,5){\makebox(0,0){\tiny{$\textnormal{denominator}(\beta)$}}}
\put(-10,12){\makebox(0,0){\tiny{$\textnormal{numerator}(\beta)$}}}
\thicklines \put(8,0){\makebox(0,0){\tiny{$\mid$}}}
\put(20,0){\makebox(0,0){\tiny{$\mid$}}}
\put(8,-2){\makebox(0,0){\tiny{1}}}
\put(-2,8){\makebox(0,0){\tiny{$1$}}}
\put(20,-2){\makebox(0,0){\tiny{$N_{s1}$}}}
\put(27,-1.5){\makebox(0,0){\tiny{$\beta$}}}
\put(0,8){\makebox(0,0){\tiny{$+$}}} \thinlines
\put(-2,10){\line(1,-1){12}} \put(-2,8.8){\line(5,-2){25}}
\end{picture}
\caption{The boundaries of $\alpha$} \label{pic:1}
\end{figure}
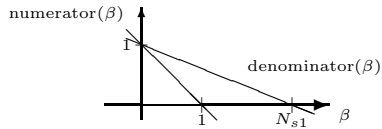
The maximum value for $\max\alpha=1$ is also conceivable from
Figure \ref{pic:1}, that makes
$P^{'}(\mathbf{E_2})|_{\ket{\eta_1}}=0$  and
$P^{'}(\mathbf{E_1})|_{\ket{\eta_2}}=\frac{1}{N_{s1}}$, which is
$2\cdot P(\mathbf{E_1})|_{\ket{\eta_2}}$. The same techniques can
be applied for $\max P(\mathbf{E_2})|_{\ket{\eta_1}}$, and it is
enough for decision if one of the two measurements or both
results $\mathbf{E_1}$ or $\mathbf{E_2}$.
\end{proof}One can make a secure decision whether $\mathbf{E_1}$ or
$\mathbf{E_2}$ or both is indicated as well as the effect of
$\mathbf{E_3}$ is reducible with repeated measurements.\par
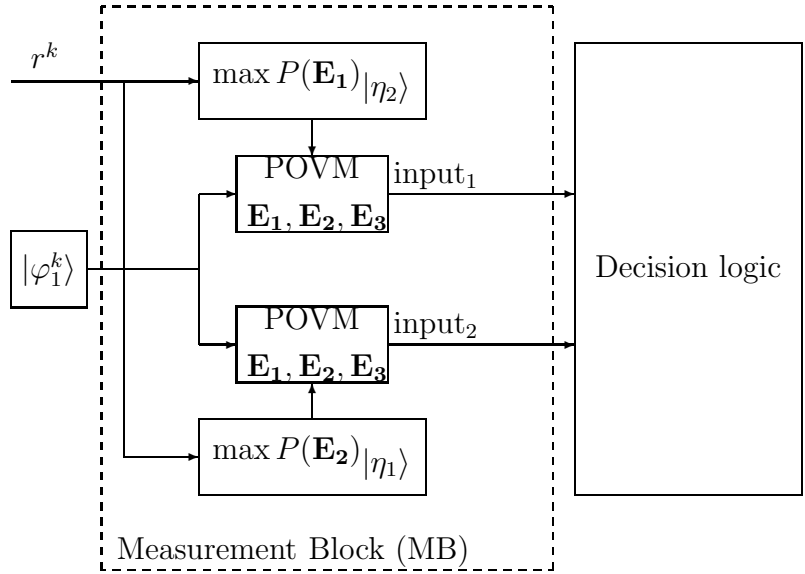
\begin{figure}[ht]
\begin{picture}(100,220)(-55,50)
\unitlength 1mm \put(10,80){\vector(1,0){25}}
\put(25,80){\line(0,-1){50}}\put(25,30){\vector(1,0){10}}
\put(10,50){\framebox(10,10){\ket{\varphi_1^k}}}%\put(0,55){\line(1,0){10}}
\put(20,55){\line(1,0){15}}
\put(35,55){\line(0,1){10}}\put(35,55){\line(0,-1){10}}\put(35,65){\vector(1,0){5}}\put(35,45){\vector(1,0){5}}
\put(35,25){\framebox(30,10){$\max
P(\mathbf{E_2})_{\ket{\eta_1}}$}}\put(35,75){\framebox(30,10){$\max
P(\mathbf{E_1})_{\ket{\eta_2}}$}}
\put(40,40){\framebox(20,10)[t]{POVM}}
\put(40,40){\framebox(20,10)[b]{
$\mathbf{E_1},\mathbf{E_2},\mathbf{E_3}$}}
\put(40,60){\framebox(20,10)[t]{POVM}}
\put(40,60){\framebox(20,10)[b]{
$\mathbf{E_1},\mathbf{E_2},\mathbf{E_3}$}}
\put(50,75){\vector(0,-1){5}}\put(50,35){\vector(0,1){5}}
\put(60,65){\vector(1,0){25}}\put(60,45){\vector(1,0){25}}
\put(13,82){\makebox(3,3){$r^k$}}
\put(65,65){\makebox(3,5){input$_1$}}
\put(65,45){\makebox(3,5){input$_2$}}
\put(85,25){\framebox(30,60){Decision logic}}
\put(22,15){\dashbox(60,75)}\put(46,10){\makebox(3,15){Measurement
Block (MB)}}
\end{picture}
\caption{Measurement block} \label{pic:2}
\end{figure}

\begin{figure}[ht]
\begin{picture}(100,200)(-70,0)
\unitlength 1mm
% boxes
\put(10,50){\framebox(10,10){\ket{\varphi_1^k}}}
\put(10,15){\framebox(10,10){\ket{\varphi_0^k}}}
\put(35,60){\framebox(10,5){MB$_1^k$}}
\put(35,52.5){\framebox(10,5){MB$_1^k$}}
\put(35,37.5){\framebox(10,5){MB$_1^k$}}
\put(37.5,43.5){\makebox(5,10){\vdots}}
\put(55,36){\framebox(22,30){Decision I.}}

\put(35,25){\framebox(10,5){MB$_0^k$}}
\put(35,17.5){\framebox(10,5){MB$_0^k$}}
\put(35,2.5){\framebox(10,5){MB$_0^k$}}
\put(37.5,8.5){\makebox(5,10){\vdots}}
\put(55,1){\framebox(22,30){Decision II.}}
\put(80,0){\framebox(17.5,67.5){Selection}}
\put(99,34){\makebox(5,3){Out}} \put(13,37){\makebox(3,3){$r^k$}}
% lines
\put(20,55){\vector(1,0){10}}\put(30,55){\vector(1,0){5}}\put(30,62.5){\vector(1,0){5}}\put(30,40){\vector(1,0){5}}
\put(30,55){\line(0,1){7.5}}\put(30,55){\line(0,-1){15}}
\put(45,61.6){\vector(1,0){10}}\put(45,54.1){\vector(1,0){10}}\put(45,39.1){\vector(1,0){10}}
\put(45,63.2){\vector(1,0){10}}\put(45,55.7){\vector(1,0){10}}\put(45,40.7){\vector(1,0){10}}

\put(20,20){\vector(1,0){10}}\put(30,20){\vector(1,0){5}}\put(30,27.5){\vector(1,0){5}}\put(30,5){\vector(1,0){5}}
\put(30,20){\line(0,1){7.5}}\put(30,20){\line(0,-1){15}}
\put(45,26.6){\vector(1,0){10}}\put(45,19.1){\vector(1,0){10}}\put(45,4.1){\vector(1,0){10}}
\put(45,28.2){\vector(1,0){10}}\put(45,20.7){\vector(1,0){10}}\put(45,5.7){\vector(1,0){10}}
\put(77,51){\vector(1,0){3}}\put(77,16){\vector(1,0){3}}\put(97.5,32.5){\vector(1,0){7}}
\put(10,35){\line(1,0){15}}\put(25,35){\line(0,1){29}}\put(25,35){\line(0,-1){32}}\put(25,64){\vector(1,0){10}}
\put(25,56.5){\vector(1,0){10}}\put(25,39){\vector(1,0){10}}\put(25,29){\vector(1,0){10}}\put(25,21.5){\vector(1,0){10}}
\put(25,3){\vector(1,0){10}}
\end{picture}
\caption{Receiver built from measurement blocks} \label{pic:3}
\end{figure}
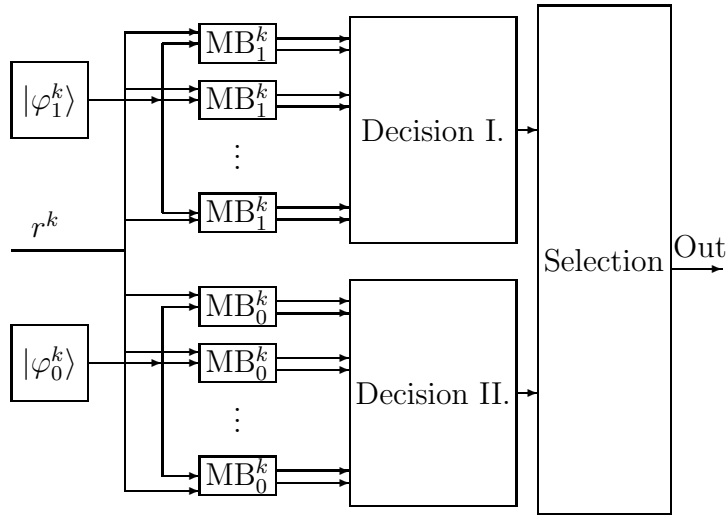

 For a
right quantum decision a Measurement Block (MB$_i^k$) can be
built up employing two POVM operations according to
\textit{Theorem 1}, as depicted in Figure \ref{pic:2}, where $k$,
$i$ refers to the user and computation base (\ref{eq:6}),
respectively. In general a detection can be made by a
\textit{Decision logic} operating on the following rule:

% (look also to \textit{Table}\ref{tab:2}) :
\begin{enumerate} \item $\mathtt{input_i}\in$
$\left\{\mathbf{E_1},\mathbf{E_2},\mathbf{E_3}\right\}$;
\item \texttt{if} $\exists $ $\mathtt{input_i} = \mathbf{E_1},$
$\mathtt{i=1,\dots,k}$ $\Rightarrow$ \texttt{out}$=\mathbf{E_1}$
\item \texttt{else} \texttt{if} $ \exists$ $ \mathtt{input_i} =
\mathbf{E_2},$ $\mathtt{i=1,\dots,k}$ $\Rightarrow$
\texttt{out}$=\mathbf{E_2}$
\item \texttt{else} \texttt{no decision}.
\end{enumerate}
In our case the \textit{Decision logics} and the \textit{Selector}
units in Figure \ref{pic:2} and \ref{pic:3} use the decision rule
table to be found in Table \ref{tab:2} according to decision rule
described above.

\par It is remarkable that in case of $Out=\mathbf{E_3}$ the
measurement block can be fed back i.e. the operation can be
repeated, which is normally equivalent to switch such a block
serial.

\begin{table}[ht]\caption{Measurement block decision rule table} \label{tab:2}
\begin{center}
\begin{tabular}[t]{|c|c|c|l|} \hline
Input 1 & Input 2 & Decision & Observation
\\\hline $\mathbf{E_1}$ & $\mathbf{E_1}$ & $\mathbf{E_1}$ & the received symbol is surely in $\ket{\varphi_1^k}$  \\ $\mathbf{E_1}$ & $\mathbf{E_3}$ & $\mathbf{E_1}$ & the received symbol is surely in $\ket{\varphi_1^k}$\\
$\mathbf{E_2}$ & $\mathbf{E_2}$ & $\mathbf{E_2}$& the received symbol is surely not in $\ket{\varphi_1^k}$\\
$\mathbf{E_2}$ & $\mathbf{E_3}$ & $\mathbf{E_2}$& the received symbol is surely not in $\ket{\varphi_1^k}$\\
$\mathbf{E_3}$ & $\mathbf{E_3}$ & $\mathbf{E_3}$& no decision\\
\hline
\end{tabular}
\end{center}
\end{table}

\vspace{1cm}

\section{Conclusions}\label{sec:4}
\par In this paper we presented a quantum computation based
multi-user detector algorithm, which involves the Positive
Operation Valued Measurement. The new method utilizes one of the
possible future receiver technologies of 3G and 4G mobile systems,
the so called quantum assisted computing. QMUD provides optimal
detection in finite time and complexity when classical methods
can achieve only suboptimal solutions. Our task is in the future
to examine and underline the in this paper given theoretical
results with some simulations.

\section*{Appendix 1: POVM-measurement}
POVM is a common used type of measurement, which provides a
secure decision, however, does not care about the state after the
measurement. The probability, notable as
$p(m)=\bra{\varphi}\underbrace{\adj{\mathbf{M_m}}\mathbf{M_m}}_{\mathbf{E_m}}\ket{\varphi}$,
where $\mathbf{E_m}$ is positive definite i.e.
$\bra{\varphi}\mathbf{E_m}\ket{\varphi}\geq 0$ and
$\sum_m\mathbf{E_m}=1$ must be satisfied. One can construct a
POVM with three elements/bases
$\{\mathbf{E_1},\mathbf{E_2},\mathbf{E_3}\}$ in such a way, that
in case of $\mathbf{E_1}$ or $\mathbf{E_2}$ unambiguous decision
is possible between two occurrence. If $\mathbf{E_3}$ is
indicated we can not decide, however, in worst case the error
correction is handled in a higher layer protocol.

\end{document}